\documentclass{article}
\usepackage{amssymb}

\begin{document}

\title{Instability of fermions with respect to topology fluctuations}
\author{A.A. Kirillov, E.P. Savelova \\
\emph{Branch of Uljanovsk State University in Dimitrovgrad, }\\
\emph{Dimitrova str 4.,} \emph{Dimitrovgrad, 433507, Russia} }
\date{}
\maketitle

\begin{abstract}
We extend our sum over topologies formula to fermions. We show that
fermionic fields display an instability with respect to topology
fluctuations. We present some phenomenological arguments for a modification
of the action in the case of fermions and discuss possible applications.
\end{abstract}

\section{Introduction}

As it was recently demonstrated in Ref. \cite{KS08}, when we take into
account for spacetime foam effects, all divergencies and inconsistencies in
quantum field theories do disappear (in the full agreement with that was
expected \cite{wheeler}). This remarkable feature however concerns only the
bosonic sector of the field theory. When we try to use the same arguments
(i.e., the same technique to account for the spacetime foam) in the case of
fermions, we find that inconsistencies only do increase. In the present
paper we analyse this feature and show the way to overcome such a difficulty.

It turns out that fermionic degrees of freedom are unstable with respect to
topology fluctuations. In the first place such an instability means\ only
that there should take place a phase transition which makes fermions to be
stable with respect to topology changes. Yet, we do not know the exact
mechanism (and the exact theory) of the stabilization and the particular
topology which makes fermions to be stable. However, we point out that this
instability is actually not a new result. It was first found by Banks et.
al. in Ref. \cite{Banks} where it was shown that wormholes drive the free
fermion mass toward the cut-off scale. If the cut-off scale is absent, then
the mass goes to infinity and therefore fermions cannot propagate. In our
picture this corresponds to the situation when the cut-off function itself
tends to zero $N\left( p\right) \rightarrow 0$ on the mass shell and
therefore fermions do not propagate. The equivalence of these two statements
is seen directly from the definition of the true Green function which has
the form $G\left( p\right) =N\left( p\right) /\left( \gamma p-m\right) $ and
which disappears in both cases as $m\rightarrow \infty $ and as $N\left(
p\right) \rightarrow 0$.

Let us recall first our sum over topologies formular \cite{KS08}. We use the
euclidean formulation, e.g., see Ref. \cite{Eu}. Then the euclidean path
integral \ for the expectation value of an observable $B$ is
\begin{equation}
\left\langle B\right\rangle =\frac{\sum Be^{-S}}{\sum e^{-S}}  \label{z1}
\end{equation}%
where $S$ is the euclidean action and the sum is taken over all field
configurations and all topologies of the euclidean spacetime. The path
integral is then taken in the two steps. First, we integrate over all field
configurations keeping a specific class of topologies fixed and then sum
over different topological classes, so that the partition function can be
presented as%
\begin{equation}
Z=\sum e^{-S}=\sum_{N}e^{-S_{eff}}  \label{z}
\end{equation}%
where $S_{eff}$ is an independent effective action for each particular
topological class and $N$ defines the topological class.

An arbitrary topology of space can be accounted for by the bias of sources
\cite{K06}. Indeed, in considering very small scales in particle physics we
use an extrapolation of spatial relationships which are well-tested only at
laboratory scales. Therefore, if the topological structure of the actual
Universe does not match properly that of the extrapolated coordinate space
we naturally should observe some discrepancy. It is convenient to describe
such a discrepancy as follows.

Consider the euclidean coordinate space $R^{4}$ and let $H$ be the Hilbert
space for a free particle which moves in $R^{4}$ (i.e., $H$ is merely the
space of functions on $R^{4}$). Let $\left\{ g_{k}\left( x\right) \right\} $
be an arbitrary basis in $H$. Physically, the basis represents a set of
eigenvectors for a complete set of observables. In our case we can consider
a scalar (without the spin) particle and use the coordinate representation,
i.e., $g_{k}\left( x\right) =\delta \left( x_{k}-x\right) $ is the set of
eigenvectors for the position operator $\widehat{X}g_{k}=x_{k}g_{k}$. The
basis is supposed to be normalized $(g_{k},g_{p})=\delta _{kp}$ and complete
$\sum g_{k}^{\ast }\left( x\right) g_{k}\left( x^{\prime }\right) =\delta
\left( x-x^{\prime }\right) $, where $x$, $x^{\prime }\in R^{4}$.

It is important that in particle physics the actual physical space $V_{phys}$
admits an embedding into the coordinate (extrapolated) space\footnote{%
We recall that when extrapolating to extremely large scales (in
astrophysics) we should consider the universal covering on which such a
simple feature $V_{phys}\subset R^{4}$ disappears. See for detail Ref.\cite%
{KS08}.} i.e., $V_{phys}\subset R^{4}$. However in general there always
exists some discrepancy between the actual space $V_{phys}$ and the
coordinate space $R^{4}$. One may easily imagine some irregular distribution
for $V_{phys}$ $\subset $ $R^{4}$ (i.e., $V_{phys}$ includes also "voids").
In terms of the scalar particle this means that some eigenvalues $x_{k}$ for
the position operator are absent. This also means that some states in $H$
cannot be physically realized for actual physical particles and fields and,
therefore, we have to restrict the space of states $H$ onto the space of
physically admissible states $H_{phys}=\widehat{P}H$, where $\widehat{P}=$ $(%
\widehat{P})^{2}$ is a projection operator. In the basis of eigenvectors the
projection operator $\widehat{P}$ takes the diagonal form $(f_{j},\widehat{P}%
f_{k})=P_{jk}=N_{k}\delta _{jk}$ with eigenvalues $N_{k}=0,1$. Thus, an
arbitrary (physically realizable) field is biased and can be presented as $%
\psi _{phys}$ $=$ $\widehat{P}^{1/2}$ $\psi $ $=$ $\sum \sqrt{N_{k}}%
a_{k}f_{k}\left( x\right) $. We see that topological structure of the space $%
V_{phys}$ is one-to-one encoded by the bias (projection) operator $\widehat{P%
}$ which can be described by its kernel, i.e., by a two point function $%
N\left( x,x^{\prime }\right) $ (which in general represents a generalized
function or a distribution). In particular, all physical observables acquire
the structure $\widehat{O}_{phys}=\widehat{P}^{1/2}~\widehat{O}~\widehat{P}%
^{1/2}$, while the physical space $V_{phys}$ of the system represents the
space of eigenvalues $x_{k}\in V_{phys}$ of the biased position operator of
the scalar particle $\widehat{X}_{phys}=\widehat{P}^{1/2}~\widehat{X}~%
\widehat{P}^{1/2}$. We also point out that from the point of view of the
mathematical coordinate space (i.e., $R^{4}$) the space $H_{phys}$ is not
complete, i.e., $\sum N_{k}f_{k}^{\ast }\left( x\right) f_{k}\left(
x^{\prime }\right) $ $=$ $N(x,x^{\prime })$ $=$ $\widehat{P}^{1/2}\delta
\left( x-x^{\prime }\right) \widehat{P}^{1/2}~$\ $\neq $ $\delta \left(
x-x^{\prime }\right) $. Thus, we see that the function $N(x,x^{\prime })$
replaces the delta function (i.e., the standard unit operator).

Consider now the Green function\footnote{%
We recall that the complete true Green functions are defined by (\ref{z1})
as $G( x,y) =<\phi ( x) \phi (y)>$ $=$ $<0|T\phi ( x) \phi ( y)|0> $.} for a
scalar wave equation in $R^{4}$
\[
\left( -\square _{x}+m^{2}\right) G\left( x,y\right) =4\pi \delta \left(
x-y\right) .
\]%
When we consider the actual physical space $V_{phys}$ $\subset $ $R^{4}$
this equation transforms as follows%
\begin{equation}
\left( -\square _{x}+m^{2}\right) G\left( x,y\right) =4\pi N\left(
x,y\right) ,  \label{gr}
\end{equation}%
where $N\left( x,y\right) $ is the bias (or projection) operator introduced
above. Let us return to the path integral (\ref{z}). Consider a particular
virtual topology of space which is one-to-one defined by specifying $N\left(
x,y\right) $. It is clear that due to symmetries of $R^{4}$ the action $S$
in (\ref{z1})-(\ref{z}) has the same value for all physical spaces which can
be obtained by rotations and transitions of the coordinate system in $R^{4}$%
. Thus, upon averaging out over possible orientations and transitions the
bias acquires always the structure $\widetilde{N}\left( x,y\right) =N\left(
\left\vert x-y\right\vert \right) $ and for the Green function we find%
\begin{equation}
G\left( x-y\right) =\int \frac{d^{4}k}{\left( 2\pi \right) ^{4}}\frac{%
N\left( k\right) }{k^{2}+m^{2}}\exp \{ik\left( x-y\right) \},  \label{gr2}
\end{equation}%
where $N\left( k\right) $ is the Fourier transform for the bias $\widetilde{N%
}( x)$. It is important that in the case of homogeneous and isotropic
background space the Green functions for fermions can be found directly from
the scalar Green function%
\begin{equation}
S_{F}\left( x-y\right) =-\left( \gamma \widehat{p}+m\right) G\left(
x-y\right)  \label{gr3}
\end{equation}%
where $\widehat{p}^{\alpha }=-i\partial ^{\alpha }$. We recall that in the
euclidean space the Dirac matrixes have the property $\{\gamma _{\alpha
}\gamma _{\beta }\}=-2\delta _{\alpha \beta }$ and therefore this Green
function obeys the equation%
\[
\left( \gamma \widehat{p}-m\right) S_{F}\left( x,y\right) =4\pi N\left(
x-y\right) .
\]

We point out that upon averaging out over orientations and transitions the
same bias $N\left( k\right) $ describes already not a particular topology
but rather a particular class of equivalent topologies which are denoted by $%
N$ in the sum (\ref{z}). Moreover, if we consider every particular
topological class, then the basic property of the projection operator $%
\widehat{P}=$ $(\widehat{P})^{2}$ (which distinguishes a particular
topologic structure) transforms into $(\tilde{P})^{2}\leq \tilde{P}$ and
therefore eigenvalues $N_{k}=0,1$ transforms in general into numbers $%
N\left( k\right) \leq 1$. However we replace further the sum over
topological classes \ (i.e., over all possible functions $N\left( k\right)
\leq 1$) with an equivalent problem that is the sum over muti-valued fields
with $N_{k}$ having the values $N_{k}=0,1$. We recall that multi-valued
fields give not more, than a specific representation for the generalized
statistics (or the Green statistics \cite{G53}) see for details Ref. \cite%
{KS08} and references therein.

Let us return to the path integral (\ref{z}). The homogeneous and isotropic
character of the background (i.e., of the coordinate) space $R^{4}$ leads to
the fact that the multi-valued character of fields is more convenient to
describe in the Fourier representation ($\phi =\frac{1}{\left( 2\pi \right)
^{2}}\int d^{4}k\phi _{k}e^{ikx}$) that is to replace a single-valued field $%
\phi _{k}$ with a set of fields $\phi _{k}^{j}$ where $j=1,2,...N\left(
k\right) $, while the bias function $N\left( k\right) $ acquires the meaning
of the number of such fields. Then the euclidean action for a field of an
arbitrary spin which in the standard picture (in the Planckian units where $%
M_{pl}=1$) takes the form%
\begin{equation}
S=\int \left[ Tr\left( \phi ^{\ast }\widehat{A}\phi \right) +V\left( \phi
\right) \right] d^{4}x,  \label{ac}
\end{equation}%
transforms into $S_{0}+S_{int}\left( \phi \right) $. Here the linear part of
the action takes the structure
\begin{equation}
S_{0}=L^{4}\int \sum_{j=1}^{N\left( k\right) }Tr\left( \phi _{k}^{j\ast
}A\left( k\right) \phi _{k}^{j}\right) \frac{d^{4}k}{\left( 2\pi \right) ^{4}%
},  \label{act}
\end{equation}%
where $A\left( k\right) =k^{2}+m^{2}$ in the case of scalar particles and $%
A\left( k\right) =\gamma k-m$ in the case of fermions. The sign $Tr$ denotes
the trace over all additional components of the field $\phi $. In the path
integral the non-linear term $S_{int}\left( \phi \right) $ is accounted for
by perturbations.

We recall that in this expression the values of the number of fields $%
N\left( k\right) $ depend on scales under consideration and, therefore, the
result for the mean cutoff function depends on the choice of the
continuation used. As it was explained previously in Ref. \cite{KS08} in
astrophysical problems we use the universal covering and the number of
fields takes values $N\left( k\right) $ $=$ $0,1,2,...$, while in particle
physics discussed previously the number of fields can take only two possible
values $N\left( k\right) =0,1$.

The physical sense has only the sum of fields ($\widetilde{\phi }%
_{k}=\sum_{j=1}^{N\left( k\right) }\phi _{k}^{j}$), and therefore the
generating functional is taken as (e.g., see the standard books Ref. \cite%
{TB})%
\[
Z\left[ J\right] =\exp \left\{ -S_{int}\left( \frac{\delta }{\delta J}%
\right) \right\} \sum_{N\left( k\right) }\widetilde{Z}\left[ J,N\right] ,
\]%
where%
\begin{eqnarray}
\widetilde{Z}\left[ J,N\right] &=&\int D\left[ \phi \right] D\left[ \phi
^{\ast }\right] \exp \left\{ -S_{0}\left( \phi \right) +L^{4}\int J^{\ast
}\left( k\right) \widetilde{\phi }_{k}d^{4}k\right\}  \nonumber \\
&=&\widetilde{Z}\left[ N\right] \exp \left\{ L^{4}\int Tr\left[ J^{\ast
}\left( k\right) A^{-1}\left( k\right) J\left( k\right) \right] \frac{%
N\left( k\right) d^{4}k}{\left( 2\pi \right) ^{4}}\right\}  \label{gf}
\end{eqnarray}%
and for $\widetilde{Z}\left[ N\right] $ we have (the sign $+$ stands for
fermions, while minus stands for bosons)%
\begin{equation}
\widetilde{Z}\left[ N\right] =\exp \left\{ \pm L^{4}\int N\left( k\right)
\frac{d^{4}k}{\left( 2\pi \right) ^{4}}\ln \frac{\det A\left( k\right) }{\pi
}\right\} .  \label{zf}
\end{equation}%
This functional has the structure%
\[
\widetilde{Z}\left[ N\right] =\prod\limits_{k}Z_{k}^{N\left( k\right) }
\]%
where $Z_{k}$ is given by the standard single-field expressions $%
Z_{k}=\left( \det A\left( k\right) /\pi \right) ^{\pm 1}$ and the sum over
possible values $N\left( k\right) =0,1$ gives%
\begin{equation}
Z=\sum_{N}\widetilde{Z}\left[ N\right] =\prod\limits_{k}\left(
\sum_{N=0,1}Z_{k}^{N\left( k\right) }\right) =\prod\limits_{k}\left(
1+Z_{k}\right) ,  \label{prt}
\end{equation}%
while for the mean cutoff we find from (\ref{z1})%
\begin{equation}
\overline{N}\left( k\right) =\frac{Z_{k}}{\left( 1+Z_{k}\right) }.
\label{cutoff}
\end{equation}%
This expression straightforwardly generalizes on an arbitrary set of fields
which gives
\begin{equation}
\ln Z_{k}=\frac{1}{2}\sum_{\alpha \in F}\ln \left( \frac{k^{2}+m_{\alpha
,F}^{2}}{\pi }\right) -\frac{1}{2}\sum_{\alpha \in B}\ln \left( \frac{%
k^{2}+m_{\alpha ,B}^{2}}{\pi }\right) ,  \label{zb}
\end{equation}%
where $F$ and $B$ stands for fermions and bosons respectively and the sum is
taken over all fields and helicity states.

When we restrict ourself with the bosonic sector only (i.e., with the second
sum in (\ref{zb}), as in Ref. \cite{KS08}), then from (\ref{zb}) we find
that on the mas-shell\footnote{%
The mas-shell requires considering the analytic continuation to the
Minkowski space.} (as $k^{2}+m_{\alpha ,B}^{2}=0$ for, at least, any
particular particle $m_{\alpha ,B}$) $Z_{k}\rightarrow \infty $ and the
cutoff (\ref{cutoff}) reduces to $\overline{N}\left( k\right) \rightarrow 1$%
. In the limit $k\rightarrow \infty $ (at very small planckian scales) $%
Z_{k}\ll 1$ and the cutoff acquires the $\overline{N}\left( k\right) \sim
1/k^{g}\rightarrow 0$ (where $g$ is the total number of bosonic degrees of
freedom). All these features disappear when we take into account for the
fermionic sector. Indeed, in the case of fermions every closed loop in
Feynman diagrams includes the additional multiplier $-1$ and therefore
fermions give contribution to (\ref{zb}) with the opposite sign. Therefore,
on the mas-shell for any particular fermion, as $k^{2}+m_{\alpha ,F}^{2}=0$,
we get $Z_{k}=0$ and $\overline{N}\left( k\right) =0$. Then from (\ref{gr2}%
), (\ref{gr3}) we find that $S_{F}\left( k\right) $ does not contain poles
(singularities ) and therefore such fermions cannot propagate. By other
words dynamics in the fermion sector disappears. Moreover, analogously the
functions $Z_{k}$, $\overline{N}\left( k\right) $ acquire a pathologic
behavior at very small scales $Z_{k}\sim k^{F-B}$ as $k\rightarrow \infty $
whose behavior depends now on the difference between the number of fermionic
and bosonic degrees of freedom. In particular, in theories where the number
of fermions exceeds that of bosons $F-B>0$ and therefore the cutoff is
absent (i.e., $\overline{N}\left( k\right) \rightarrow 1$ as $k\rightarrow
\infty $).

Origin of such an anomalous behavior is quite clear. First we recall that
the action for fermions has not a clear classical analogue (the formal
analogue is a two-level system or Grassman fields). Therefore, in the case
of fermions the standard expressions for the action or the energy density
admit an ambiguity, i.e., they admit a shift on an arbitrary function. While
topology is fixed and cannot change, such a shift renormalizes merely the
cosmological constant and gives no contribution to any observables in
particle physics (save gravitational physics). However, when topology may
change (or fluctuate) such a shift becomes an issue; for it defines the
resulting stable topology of the actual physical space.

Consider the energy for a particular field mode which in the standard
picture has the form%
\[
E=\varepsilon _{k}\left( n_{k}\pm \frac{1}{2}\right)
\]%
where the sign $+/-$ stands for bosons/fermions, $n_{k}=0,1,2,...$ in the
case of bosons, and $n_{k}=0,1$ in the case of fermions. The ground state of
the mode has the energy $E_{0}=\pm \varepsilon _{k}/2$. Topology changes
allow to remove some of such oscillators and therefore such a process
accompanies always with a decrease/increase of the total energy by the
factor $\Delta E=\pm \sum_{k}\varepsilon _{k}/2$ where the sum is taken over
those oscillators which are "removed". It is quite natural to expect that
when all modes are absent, then the field itself (and therefore the physical
space itself) is absent \cite{K99}. In this case there are no observables
related to the field at all and such a case is convenient to imagine as an
absolute vacuum (absolute ground state). Such a feature holds indeed in the
case of bosonic oscillators and it is tempting to suppose that creation of
the physical space requires to spent some energy. However, we see that in
the case of fermionic systems the minimum is always reached when all
oscillators are present, i.e., the bias has the form $N_{k}=1$, while the
true vacuum (i.e., when the field is totally absent $N_{k}=0$) has bigger
energy and therefore is unstable\footnote{%
such an instability is more prominent in the astrophysics where we have to
use the universal covering \cite{KS08} and the number of fermionic modes can
be $N(k)=0,1,2,...$. Then the instability leads to the formation of the more
and more number of fermionic modes i.e., $N(k)\rightarrow \infty $. In this
case the minimum energy is $E\rightarrow -\infty$.}. All our knowledge in
physics teaches us that any unstable situation leads to some phase
transition upon which the system has to be stable.

Let us re-define the action (\ref{act}) in the form
\[
S_{0}=L^{4}\int \sum_{j=1}^{N\left( k\right) }\left[ Tr\left( \phi
_{k}^{j\ast }A\left( k\right) \phi _{k}^{j}\right) +\lambda \left( k\right) %
\right] \frac{d^{4}k}{\left( 2\pi \right) ^{4}},
\]%
where $\lambda \left( k\right) $ is yet an arbitrary function. In the case
of bosons (since we can fix the action principle by the classical limit) we
may expect that $\lambda \left( k\right) =const$, while in the case of
fermions this function requires an additional consideration. The presence of
such an additional function leads to the following modification of (\ref{zb}%
)
\[
\ln Z_{k}=\ln Z_{k}^{F}+\ln Z_{k}^{B},
\]%
where
\begin{equation}
\ln Z_{k}^{F(B)}=\sum_{\alpha \in F(B)}\left[ \lambda _{\alpha ,F(B)}\left(
k\right) \pm \frac{1}{2}\ln \left( \frac{k^{2}+m_{\alpha ,F(B)}^{2}}{\pi }%
\right) \right] .  \label{lam}
\end{equation}%
It is convenient to re-write this in a more symmetric manner as follows
\begin{equation}
\ln Z_{k}^{F(B)}=\sum_{\alpha \in F(B)}\frac{1}{2}\ln \left( \frac{\mu
_{\alpha }^{F(B)}\left( k\right) }{k^{2}+m_{\alpha ,F(B}^{2)}}\right) ,
\label{mu}
\end{equation}%
where parameters $\mu _{\alpha }^{F(B)}$ are defined by comparing (\ref{lam}%
) and (\ref{mu}).

As it was discussed in Ref. \cite{KS08} in the case of bosons the most
natural choice is $\mu _{\alpha }^{B}\left( k\right) =\mu $ (i.e., $\lambda
\left( k\right) =const$) where $\mu $ characterizes the scale of the actual
cutoff. Analogously, we may expect that upon the phase transition (i.e.,
when fermions become stable) such a parameter takes the value of the same
actual cutoff $\mu _{\alpha }^{F}\left( k\right) =\mu $. Indeed, let us
consider some coarse graining in the phase space, i.e., each value $k_{j}$
will correspond to some volume $\Delta k^{3}$, so that there will be a
sufficiently big number of states in an every coarse grained state $k_{j}$.
Then the difference between fermions and bosons should disappear (upon the
coarse graining, more than one fermion can occupy the same quantum state $%
k_{j}$) and, therefore, we may expect that fermions and bosons should give a
symmetric contribution to the cutoff function $\overline{N}\left( k\right) $%
\footnote{%
We understand that such arguments are far from being rigorous though.
However in the absence of the complete theory for topology changes, we may
use only such phenomenological arguments. At least the only rigorous
criteria here may come from experiments.}.

By other words in particle physics the cutoff function always acquires the
structure (see for details Ref. \cite{KS08})
\begin{equation}
\overline{N}\left( k\right) =\frac{\mu ^{g}}{\left( \mu ^{g}+k^{2\alpha
_{0}}\left( k^{2}+m_{1}^{2}\right) ^{\alpha _{1}}\cdots \left(
k^{2}+m_{n}^{2}\right) ^{\alpha _{n}}\right) }  \label{cut}
\end{equation}%
where $\mu $ is the cutoff scale, $g=\sum 2\alpha _{n}$ is the total number
of fields (both fermions and bosons) we have to retain, and the cutoff
parameter $\mu $ can be defined via the total (observational) cosmological
constant term.

\end{document}